\documentclass{iau} 
\usepackage{graphicx}

\title[JD 11.~~Metal-poor and metal-rich GCs in E-MOSAICS] %% give here short title %%
{The origin of metal-poor and metal-rich globular clusters in E-MOSAICS}

\author[Marta Reina-Campos]   %% give here short author list %%
{Marta Reina-Campos}

\affiliation{Astronomisches Rechen-Institut, Zentrum f\"{u}r Astronomie der Universit\"{a}t Heidelberg, M\"{o}nchhofstra\ss e 12-14, 69120 Heidelberg, Germany \\ email: {\tt reina.campos@uni-heidelberg.de} \\[\affilskip]}

\pubyear{2019}
\volume{351}  
\jname{Star Clusters: From the Milky Way to the Early Universe}
\editors{A. Bragaglia, M.B. Davies, A. Sills \& E. Vesperini, eds.}
\begin{document}

\maketitle
%. CONTINUE EDITING FROM HERE

\begin{abstract}
It has been a long-standing open question why observed globular cluster (GC) populations of different metallicities differ in their ages and spatial distributions, with metal-poor GCs being the older and radially more extended of the two. We use the suite of 25 Milky Way-mass cosmological zoom-in simulations from the E-MOSAICS project, which self-consistently model the formation and evolution of stellar clusters and their host galaxies, to understand the properties of observed GC populations. We find that the different ages and spatial distributions of metal-poor and metal-rich GCs are the result of regular cluster formation at high redshift in the context of hierarchical galaxy assembly. We also find that metallicity on its own is not a good tracer of accretion, and other properties, such as kinematics, need to be considered.
\end{abstract}

\firstsection % if your document starts with a section,
\section{Introduction}

Due to its close proximity, the Galactic globular cluster (GC) system is the most studied cluster population in the Local Universe and most of the theoretical effort in the literature has been devoted into understanding its properties. However, over the past 60 years, our picture of GC systems has evolved tremendously as more observations of extragalactic GC systems are becoming available. It has been observed that GC systems host a variety of properties regarding their masses, ages, metallicities, kinematics and positions (e.g. \cite[Brodie \& Strader 2006]{brodie06}, \cite[Portegies Zwart, McMillan \& Gieles 2010]{portegieszwart10}). Compared to other extragalactic systems, the Galactic GC system is relatively older and more metal-poor, which suggests that it assembled earlier in cosmic history (e.g. \cite[Haywood et al. 2013]{haywood13}, \cite[Snaith et al. 2014, 2015]{snaith14,snaith15}, \cite[Mackereth et al. 2018]{mackereth18a}, \cite[Kruijssen et al. 2019b]{kruijssen19b}).

Despite the observed variety of GC populations, some of the properties appear to be relatively universal when dividing the GC populations by metallicity. Firstly, GC ages are correlated with metallicity, with the most metal-poor GCs being the oldest (e.g. \cite[Forbes \& Bridges 2010]{forbes10}). The existence of several of such GC age-metallicity relations within individual galaxies has been suggested to correspond to relics of different accretion events (\cite[Kruijssen et al. 2019a]{kruijssen19a}). Over the past year, these GC age-metallicity relations along with their kinematics from Gaia DR2 have been used to identify accretion events in our Galaxy (\cite[Kruijssen et al. 2019b]{kruijssen19b}, \cite[Massari, Koppelman \& Helmi 2019]{massari19}).

Secondly, GC populations of different metallicities have been observed to have different spatial distributions, with the metal-poor subpopulation being more radially extended than metal-rich GCs (e.g. \cite[Brodie \& Strader 2006]{brodie06}). This has been suggested to indicate that each metallicity subpopulation formed in a different environment; i.e. the extended population of metal-poor GCs formed in accreted satellite galaxies, whereas the more centrally concentrated population of metal-rich objects formed in-situ in their host galaxy.

The origin of the different extent of the GC subpopulations, alongside their age differences and kinematics, are points of debate among the community. Some authors suggest that the different properties are signatures of two different formation channels for each metallicity subpopulation (e.g. \cite[Griffen et al. 2010]{griffen10}, \cite[Trenti, Padoan \& Jimenez 2015]{trenti15}), whereas other authors suggest that these different properties are the natural outcome of cluster formation in a context of hierarchical galaxy assembly (e.g.~\cite[Kravtsov \& Gnedin 2005]{kravtsov05}, \cite[Kruijssen 2015]{kruijssen15}).

In order to test the latter of these hypotheses, we use the 25 present-day Milky Way-mass cosmological zoom-in simulations from the E-MOSAICS project (\cite[Pfeffer et al. 2018]{pfeffer18}, \cite[Kruijssen et al. 2019a]{kruijssen19a}). These simulations self-consistently model the co-formation and evolution of stellar clusters and their host galaxies, and test the idea that the massive clusters identified nowadays as GCs are the product of cluster formation at high redshift that survive to the present day. In the simulations, we model stellar clusters as a subgrid component of the stellar population, and their formation and evolution depends on the local environment. With this formalism, the simulated cluster populations have been found to reproduce many key observed properties of observed cluster systems in the Local Universe (e.g. \cite[Usher et al. 2018]{usher18}, \cite[Kruijssen et al. 2019a,b]{kruijssen19a,kruijssen19b}, \cite[Pfeffer et al. 2019]{pfeffer19b}). For more details, we refer the readers to \cite[Pfeffer et al. (2018)]{pfeffer18} and \cite[Kruijssen et al. (2019a)]{kruijssen19a}.

\section{Properties of metal-poor and metal-rich globular clusters}

Using the 25 present-day Milky Way-mass galaxies from the E-MOSAICS simulations, we study different properties of the massive stellar cluster populations, for different metallicity subpopulations. 

\begin{figure}[b]
% \vspace*{-2.0 cm}
\begin{center}
 \includegraphics[width=5in]{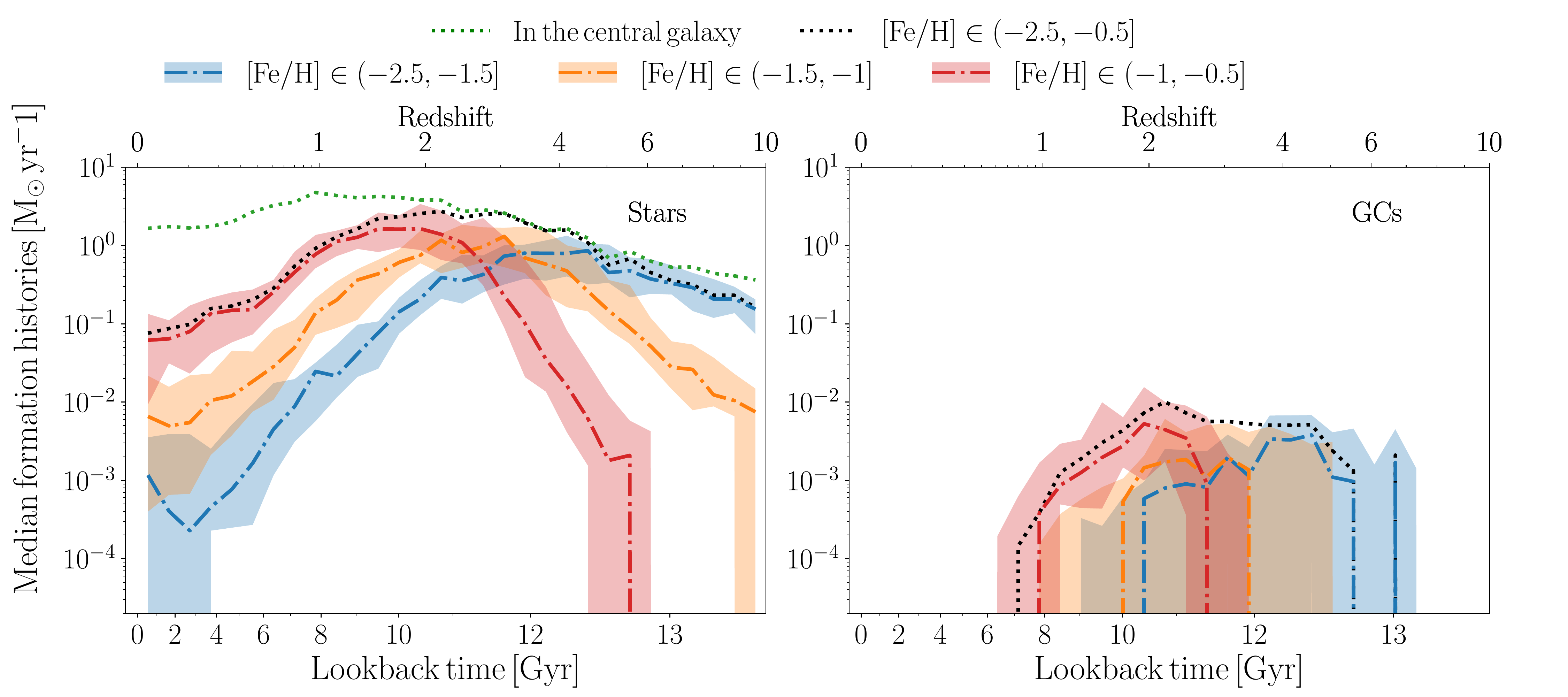} 
% \vspace*{-1.0 cm}
 \caption{Median formation histories of stars and massive stellar clusters ($M>10^5~M_{\odot}$) among the 25 present-day Milky Way-mass galaxies of the E-MOSAICS simulations for different metallicity bins. Figure taken from \cite[Reina-Campos et al. 2019]{reina-campos19}.}
 \label{fig1}
\end{center}
\end{figure}

We first consider the different formation epochs of GCs as a function of metallicity. For that, we identify GCs in our simulations as massive stellar clusters ($M>10^5~M_{\odot}$) with metallicities in the range $\rm [Fe/H]\in(-2.5, -0.5]$. With this definition, in Fig.~\ref{fig1} we show the median formation histories of stars and GCs among our suite of galaxies for different metallicity ranges. We find that star and cluster formation are continuous processes; i.e. as galaxies collapse and grow, the gas in them, from which both stars and cluster form, gets enriched and more metal-rich objects can form.

Comparing to the Galactic GC system, we find that the GC ages measured from our median GC formation histories are consistent with the median observed ages (\cite[Reina-Campos et al. 2019]{reina-campos19}). In particular, we find that the Galactic metal-poor GCs are particularly old relative to our cluster populations, suggesting an early formation and assembly of our Galaxy (e.g. \cite[Haywood et al. 2013]{haywood13}, \cite[Snaith et al. 2014, 2015]{snaith14,snaith15}, \cite[Mackereth et al. 2018]{mackereth18a}, \cite[Kruijssen et al. 2019b]{kruijssen19b}).

Secondly, we investigate the spatial distributions of the GC populations in our suite of simulations, and test the hypothesis that metal-poor GCs are the accreted subpopulation. In order to employ a definition similar to typical observational GC selection functions, we refine our GC definition to only include those older than $6~\rm Gyr$. 

\begin{figure}[b]
% \vspace*{-2.0 cm}
\begin{center}
 \includegraphics[width=5in]{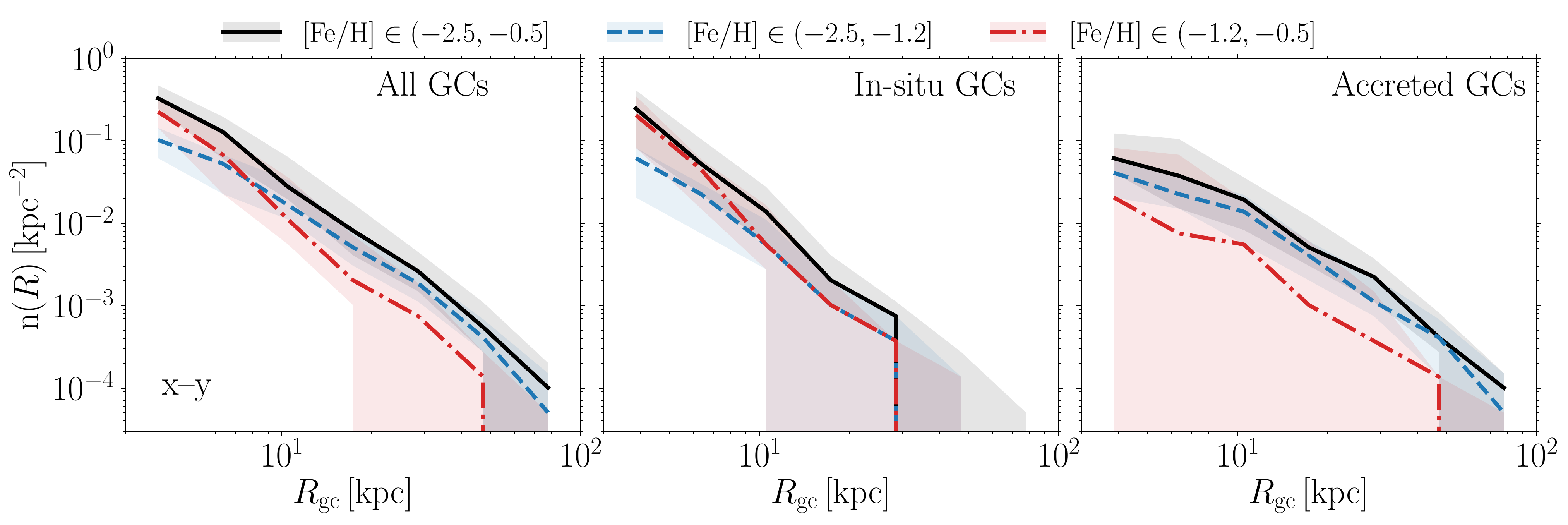} 
% \vspace*{-1.0 cm}
 \caption{Median projected number density profiles of massive stellar clusters ($M>10^5~M_{\odot}$) among the 25 Milky Way-mass galaxies of the E-MOSAICS simulations for different metallicity bins. From left to right, we show the profiles of all clusters in the central galaxies, of those that formed in the central galaxy (i.e. \emph{in-situ}) and of those that formed in \emph{accreted} satellite galaxies. Figure taken from \cite[Reina-Campos et al. 2019]{reina-campos19b}.}
   \label{fig2}
\end{center}
\end{figure}

In Fig.~\ref{fig2}, we show the median projected number density profiles of GCs among our suite of 25 present-day Milky Way-mass galaxies for different metallicity bins. We find that the simulated cluster populations reproduce the observed difference in spatial extent of each metallicity subpopulation, with the metal-poor objects being more radially extended than metal-rich GCs. 

We investigate the hypothesis that metal-poor GCs are the accreted component by dividing our GC populations between those that formed in-situ in their host galaxy and those that formed in an accreted satellite galaxy. We find that the in-situ population of GCs is dominated by the metal-rich subpopulation in the inner regions, but both metallicity subpopulations contribute equally in the outskirts. In the case of the accreted subpopulation of GCs, we find that is dominated by metal-poor objects at all radii, but there is a non-negligible component of metal-rich objects in the radial range studied. This shows that it is an oversimplification to interpret metal-poor GCs as accreted and metal-rich GCs as in-situ.

These results are consistent with the picture that GCs formed as the products of regular cluster formation during hierarchical galaxy assembly (e.g.~\cite[Kravtsov \& Gnedin 2005]{kravtsov05}, \cite[Kruijssen 2015]{kruijssen15}). On the one hand, and as represented in Fig. 1, cluster formation is a continuous process, so the in-situ component is composed of more metal-rich objects over time. On the other hand, due to the mass-metallicity relation observed for galaxies (e.g. \cite[Maiolino \& Manucci 2019]{maiolino19}), the more massive satellites bring in more metal-rich cluster populations than the less massive satellites. With this, the accreted component of GCs in their final host galaxy will be more metal-rich if a larger proportion of massive satellites is accreted. Lastly, we see that metallicity on its own is not a good tracer of accretion, and other properties, such as kinematics or chemical abundances, need to be considered.

To summarise, we use the suite of 25 present-day Milky Way-mass galaxies from the E-MOSAICS simulations to study the different properties of GC systems as a function of metallicity. We find that the different formation epochs and spatial distributions observed for the metal-poor and metal-rich GC populations can be well understood in a scenario in which the products of regular cluster formation at high redshift are redistributed during hierarchical galaxy assembly, and eventually survive as the GC populations seen at the present day.

\section*{Acknowledgements}
MRC is supported by a PhD Fellowship from the IMPRS-HD. MRC acknowledges funding from the European Research Council (ERC-StG-714907, MUSTANG). This work used the DiRAC Data Centric system at Durham University, operated by the Institute for Computational Cosmology on behalf of the STFC DiRAC HPC Facility. This equipment was funded by BIS National E-infrastructure capital grant ST/K00042X/1, STFC capital grants ST/H008519/1 and ST/K00087X/1, STFC DiRAC Operations grant ST/K003267/1 and Durham University. DiRAC is part of the National E-Infrastructure. The work also made use of high performance computing facilities at Liverpool John Moores University, partly funded by the Royal Society and LJMU’s Faculty of Engineering and Technology.

\end{document}